\documentclass{article}[12pt]

\usepackage{amsthm,amsfonts,amsmath,amssymb,array,graphicx,graphics,tabularx}
\usepackage[mathscr]{eucal}
\usepackage{times}

\newcommand{\I}{\mathrm{i}}
\newcommand{\E}{\mathrm{e}}

\newcommand{\im}{\mathop{\mathrm{Im}}}
\numberwithin{equation}{section}
\theoremstyle{plain}

\newtheorem{prop}{Proposition}[section]

\theoremstyle{definition}

\theoremstyle{remark}

\allowdisplaybreaks[4]

\title{A ray mode parabolic equation for shallow water acoustics propagation problems}

\author{M.~Yu.~Trofimov, A.D.~Zakharenko}
\date{Il'ichev Pacific Ocenological Institute FEBRAS and Far Eastern
Federal University, Vladivostok, Russia}

\begin{document}

\maketitle

\begin{abstract}
  Ray mode parabolic equations which are suitable for shallow water acoustics propagation problems are derived by the multiple-scale method.
\end{abstract}

\section{Introduction}

After the work of R.~Burridge and H.~Weinberg \cite{Burridge} it was natural to derive
mode parabolic equation. It was done by M.~Collins \cite{collins} in the framework of operator approach.
Later the strict approach to this problem (method of multiple-scale
expansions) was developed in our works \cite{trofimov1,trofimov2}.
\par
The practice of coastal underwater acoustics requires now more flexible approach to computation of acoustic fields;
the reflection of sound from steep shoare should be included in the propagation model. We believe that this can be done
by the ray mode parabolic equations.
\par
History of ray parabolic equations is connected mostly with Babich's school \cite{babich}. The derivation of such equations was done
in the `ray centered' coordinate system, and we show  that this choice of coordinate system is the genesis
of quadratic term in potential of resulting patabolic equation.
\par
In this work we make an attempt to derive a ray mode parabolic equation in the ray coordinates by the multiple-scale method.
We show that by introducing correctky scaled variables the equations of such a type are produced almost automatically.
As a byproduct of our work we present the one-mode Helmholtz equation, to which the standard technique of Babich's school can be applied.
In this way we obtain also the ray mode parabolic equation in ray centered coordinates.

\section{Parabolic equation in ray coordinates}
We consider the propagation of time-harmonic sound in the three-dimen\-si\-onal waveguide
$\Omega=\{(x,y,z)| 0 \leq x\leq \infty, -\infty \leq y\leq \infty,  -H\leq z \leq 0 \}$ ($z$-axis is directed upward),
described by the acoustic Helmholtz equation
\begin{equation} \label{Helm}
\left(\gamma P_x\right)_x + \left(\gamma P_y\right)_y
 + \left(\gamma P_z\right)_z + \gamma\kappa^2 P = 0\,,
\end{equation}
where $\gamma = 1/\rho$, $\rho=\rho(x,y,z)$ is the density, $\kappa$ is the wave-number. We assume the appropriate radiation conditions at infinity in $x,y$ plane, the pressure-release
boundary condition at $z=0$
\begin{equation} \label{Dir}
P=0\quad \text{at}\quad z=0\,,
\end{equation}
and rigid boundary condition $\partial u/\partial z=0$ at $z= -H$.
The parameters of medium may be discontinuous at the nonintersecting smooth interfaces  $z=h_1(x,y),\ldots,h_m(x,y)$,
where the usual interface conditions\sloppy
\begin{equation}\label{InterfCond}
\begin{split}
& P_+ = P_-\,,\\
& \gamma_+\left(\frac{\partial P}{\partial z}-h_x\frac{\partial
P}{\partial x}- h_y\frac{\partial P}{\partial y}\right)_+ =
\gamma_-\left(\frac{\partial P}{\partial z}-h_x\frac{\partial
P}{\partial x}- h_y\frac{\partial P}{\partial y}\right)_-\,,
\end{split}
\end{equation}
are imposed. Hereafter we use the denotations $f(z_0,x,y)_+=\lim_{z\downarrow z_0}f(z,x,y)$ and
$f(z_0,x,y)_-=\lim_{z\uparrow z_0}f(z,x,y)$.
\par
As will be seen below, it is sufficient to consider the case $m=1$, so we set $m=1$ and denote $h_1$ by $h$.
\par
We introduce a small parameter $\epsilon$ (the ratio
of the typical wavelength to the typical size of medium inhomogeneities), the slow
variables $X=\epsilon x$ and $Y=\epsilon y$, the fast variables $\eta=(1/\epsilon) \Theta(X,Y)$ and $\xi=(1/\sqrt{\epsilon})\Psi(X,Y)$  and postulate the following
expansions for the acoustic pressure $P$ and the parameters $\kappa^2$, $\gamma$ and $h$:
\begin{equation}\label{expans}
\begin{split}
 & P=P_0(X,Y,z,\eta,\xi)+\sqrt{\epsilon} P_{1/2}(X,Y,z,\eta,\xi)+\cdots\,,\\
 & \kappa^2 = n_0^2(X,Y,z) + \epsilon\nu(X,Y,z,\xi)\,, \\
 & \gamma  =  \gamma_0(X,Y,z) + \epsilon\gamma_1(X,Y,z,\xi)\,,\\
 & h = h_0(X,Y) + \epsilon h_1(X,Y,\xi)\,.
\end{split}
\end{equation}

To model the attenuation effects we admit $\nu$ to be complex. Namely, we take $\im\nu = 2\mu\beta n_0^2$, where
$\mu = (40\pi\log_{10}e)^{-1}$ and $\beta$ is the attenuation  in decibels per wavelength. This implies that
$\im\nu\ge 0$.
Following the generalized multiple-scale method, we replace derivatives in equation~(\ref{Helm}) by the rules
\begin{equation*}
\begin{split}
& \frac{\partial}{\partial x} \rightarrow \epsilon\left(\frac{\partial}{\partial X}+\frac{1}{\sqrt{\epsilon}}\Psi_X\frac{\partial}{\partial \xi}
+\frac{1}{\epsilon}\Theta_X\frac{\partial}{\partial \eta}\right)\,,\\
& \frac{\partial}{\partial y} \rightarrow \epsilon\left(\frac{\partial}{\partial Y}\,+\frac{1}{\sqrt{\epsilon}}\Psi_Y\frac{\partial}{\partial \xi}
+\frac{1}{\epsilon}\Theta_Y\frac{\partial}{\partial \eta}\right)\,.
\end{split}
\end{equation*}
 With the postulated expansions, the equation under consideration is
\begin{equation} \label{2}
\begin{split}
&\epsilon^2\left(\frac{\partial}{\partial X}+\frac{1}{\sqrt{\epsilon}}\Psi_X\frac{\partial}{\partial \xi}
+\frac{1}{\epsilon}\Theta_X\frac{\partial}{\partial \eta}\right)
\left((\gamma_0+\epsilon\gamma_1)\right.\\
& \quad \cdot\left.\left(\frac{\partial}{\partial X}+\frac{1}{\sqrt{\epsilon}}\Psi_X\frac{\partial}{\partial \xi}
+\frac{1}{\epsilon}\Theta_X\frac{\partial}{\partial \eta}\right)
 \cdot
 \left(\vphantom{\frac{\partial}{\partial X}}P_0+\epsilon P_1+\cdots\,,\right)\right)\\
& + \text{the same term with the $Y$-derivatives}
 + \left((\gamma_0+\epsilon\gamma_1)\left(P_{0z}+\epsilon P_{1z}+\cdots\,,\right)\right)_z\\
& + (\gamma_0+\epsilon\gamma_1)(n_0^2 + \epsilon\nu)\left(P_0+\epsilon P_1+\cdots\,,\right)=0\,.
\end{split}
\end{equation}
We put now
\begin{equation*}
\begin{split}
& P_0+\epsilon P_1+\cdots=\\
& (A_0(X,Y,z,\xi)+\epsilon A_1(X,Y,z,\xi)+\cdots\,,)\E^{\I\eta}\,.
\end{split}
\end{equation*}
Using the Taylor expansion, we can formulate the interface conditions at $h_0$ which are equivalent
to interface conditions (\ref{InterfCond}) up to $O(\epsilon)$:
\begin{equation}\label{InterfCondh01}
\left(A_{0}+\epsilon h_1 A_{0z}+\epsilon A_{1}\right)_+ =(\text{the same terms})_-\,,
\end{equation}
\begin{equation}\label{InterfCondh02}
\begin{split}
& \left((\gamma_{0}+\epsilon h_1 \gamma_{0z}+\epsilon \gamma_{1})\right. \\
& \qquad\left.\left(A_{0z}+\epsilon h_1 A_{0zz}+\epsilon A_{1z}  - \epsilon\mathrm{i}\Theta_X h_{0X}A_{0}
 - \epsilon\mathrm{i}\Theta_Y h_{0Y}A_{0}\right)\right)_+ \\
& = \left(\mbox{the same terms}\right)_-\,.
\end{split}
\end{equation}
\subsection{The problem at $O\epsilon^{0})$}
At $O\epsilon^{0})$ we obtain
\begin{equation}\label{E0}
(\gamma_0 A_{0z})_z + \gamma_0 n^2_0 A_0 - \gamma_0\left((\Theta_X)^2 + (\Theta_Y)^2\right)A_0 = 0\,,
\end{equation}
with the interface conditions of the order $\epsilon^{0}$
\begin{equation}\label{InterfaceE0}
\begin{split}
& A_{0+} = A_0{-}\,,\\
& \left(\gamma_0 A_{0z}\right)_+ =
\left(\gamma_0 A_{0z}\right)_-\quad\mbox{at}\quad z=h_0\,,
\end{split}
\end{equation}
and boundary conditions $A_0=0$ at $z=0$ and $\partial A_0/\partial x$ at $z= -H$.
We seek a solution to problem (\ref{E0}), (\ref{InterfaceE0}) in the form
\begin{equation} \label{anz0}
A_0 = B(X,Y,\xi)\phi(X,Y,z)\,.
\end{equation}
From eqs.~(\ref{E0}) and (\ref{InterfaceE0}) we obtain the following spectral problem
for $\phi$ with the spectral parameter $k^2 = (\Theta_X)^2 + (\Theta_Y)^2$
\begin{equation} \label{Spectral}
\begin{split}
& \left(\gamma_0\phi_z\right)_z + \gamma_0 n_0^2 \phi - \gamma_0 k^2 \phi=0\,,\\
& \phi(0) = 0\,,
\quad \phi_z=0\quad \text{at}\quad z= -H\,,\\
& \phi_+ = \phi_-\,,\quad \left(\gamma_0 \phi_z\right)_+ =
\left(\gamma_0 \phi_z\right)_-\quad\mbox{at}\quad z=h_0\,.
\end{split}
\end{equation}
This spectral problem, considering in the Hilbert space $L_{2,\gamma_0}[-H,0]$ with the scalar product
\begin{equation} \label{L2scalar}
(\phi,\psi) = \int_{-H}^{\,0}\gamma_0 \phi\psi\,dz\,,
\end{equation}
has countably many solutions $(k_j^2,\phi_j)$, $j=1,2,\ldots$ where the eigenfunction can be chosen as real functions.
The eigenvalues $k_j^2$ are real and have $-\infty$
as a single accumulation point.
Let $A_0=B_j(X,Y)\phi_j(X,z)$ where $\phi_j$ is a normalized eigenfunction with the corresponding eigenvalue
$k_j^2>0$ and $B_j$ is an amplitude function to be determined at the next orders of $\epsilon$.
The normalizing condition is
\begin{equation} \label{norm}
(\phi,\phi) = \int_{-H}^{\,0}\gamma_0 \phi^2\,dz=1\,,
\end{equation}

\subsection{The problem at $O\epsilon^{1/2})$}

The solvability condition of problem at $O(\epsilon^{1/2})$ is
\begin{equation} \label{1/2}
\Theta_X\Psi_X+\Theta_Y\Psi_Y=0\,,
\end{equation}
from which we coclude that
\begin{equation} \label{orth1/2}
\nabla\Theta \bot \nabla\Psi\,,
\end{equation}
and that we can take $P_{1/2}=0$.

\subsection{The problem at $O\epsilon^{1})$}

At $O(\epsilon^{1})$ we obtain
\begin{equation}\label{E1}
\begin{split}
&\left(\gamma_0A_{1z}\right)_z + \gamma_0 n_0^2 A_{1} - \gamma_0 k_j^2 A_{1}
 = -\mathrm{i}\gamma_{0X}k_j A_0 -2\mathrm{i}\gamma_{0}k_j A_{0X}\\
& -\mathrm{i}\gamma_{0}k_{jX} u_0
+ \gamma_{1}k_j^2 A_0 - \gamma_{0}(\Psi_X)^2 A_{0\xi\xi} \\
& -\text{the same terms with $Y$-derivatives}
 - \frac{\partial}{\partial z}\left(\gamma_{1} A_{0z}\right) - n_0^2\gamma_{1}A_0 -\nu\gamma_{0}A_0 \,,
\end{split}
\end{equation}
with the boundary conditions $A_1=0$ at $z=0$, $\partial A_1/\partial z=0$ at $z=-H$,
and the interface conditions at $z=h_0(X,Y)$:
\begin{equation}\label{InterfCondE1}
\begin{split}
& A_{1+}-A_{1-} = h_1(A_{0z-}-A_{0z+})\,, \\
& \gamma_{0+}A_{1z+}-\gamma_{0-}A_{1z-}
 = h_1\left(\left( (\gamma_0 A_{0z})_z\right)_-  -
\left( (\gamma_0 A_{0z})_z\right)_+\right) \\
&   + \gamma_{1-}A_{0z-}-\gamma_{1+}A_{0z+}
 -\mathrm{i}k_j h_{0X}A_0(\gamma_{0-}-\gamma_{0+})\\
& \qquad -\mathrm{i}k_j h_{0Y}A_0(\gamma_{0-}-\gamma_{0+})
\end{split}
\end{equation}
Multiplying (\ref{E1}) by $\phi_j$ and then integrating resulting equation from
$-H$ to $0$ by parts twice with the use of interface conditions (\ref{InterfCondE1}), we obtain
\begin{prop}\label{pr_MPE}
The solvability condition for the problem at $O(\epsilon^1)$ where $A_0=B_j\phi_j$ is
\begin{equation}\label{MPE}
\begin{split}
& 2\mathrm{i}(\Theta_{jX} B_{jX}+\Theta_{jY} B_{jY})
+ \mathrm{i}(\Theta_{jXX}+\Theta_{jYY}) B \\
& \qquad\qquad\qquad +
((\Psi_X)^2+(\Psi_Y)^2) B_{j\xi\xi} + \alpha_j B_j = 0\,,
\end{split}
\end{equation}
where $\alpha_j$ is given by the following formula
\begin{equation}\label{alpha}
\begin{split}
&\alpha_j = \int_{-\infty}^0 \gamma_0\nu \phi^2_j\,dz
 +
\int_{-\infty}^0 \gamma_1\left(n_0^2-k_j^2\right) \phi^2_j\,dz
 -
\int_{-\infty}^0 \gamma_1\left(\phi_{jz}\right)^2\,dz  \\
& \qquad + \left\{h_1\phi_j\left[\left((\gamma_0\phi_{jz})_z\right)_+ - \left((\gamma_0\phi_{jz})_z\right)_- \right]
\vphantom{ \left(\frac{\gamma_1}{\gamma_0}\right)_-}
\right.\\
& \qquad\qquad\left.\left.- h_1\gamma_0^2\left(\phi_{jz}\right)^2\left[\left(\frac{1}{\gamma_0}\right)_+ -
\left(\frac{1}{\gamma_0}\right)_-\right]\right\}\right|_{z=h_0}
\,.
\end{split}
\end{equation}
\end{prop}
Using spectral problem~(\ref{Spectral}) the interface terms in (\ref{alpha}) can be rewritten also as
\begin{equation*}
\begin{split}
&  \left\{h_1\phi_j^2\left[k_j^2\left(\gamma_{0+}-\gamma_{0-}\right)
- \left(n_0^2\gamma_0\right)_+ + \left(n_0^2\gamma_0\right)_-  \right]
\vphantom{ \left(\frac{\gamma_1}{\gamma_0}\right)_-}
\right.\\
& \left.\left.- h_1\gamma_0^2\left(\phi_{jz}\right)^2\left[\left(\frac{1}{\gamma_0}\right)_+ -
\left(\frac{1}{\gamma_0}\right)_-\right]\right\}\right|_{z=h_0}
\,.
\end{split}
\end{equation*}
Consider the ray equations for the Hamilton-Jacobi equation
\[
 (\Theta_X)^2 + (\Theta_Y)^2=p^2+q^2=k^2
\]
 in the form
\begin{equation}\label{ray}
\begin{split}
X_t &=\frac{p}{k}\\
Y_t &=\frac{q}{k}\\
p_t &=\frac{p}{k^2}k_X+k_X\\
q_t &=\frac{q}{k^2}k_Y+k_Y
\end{split}
\end{equation}
We have $(X_t)^2+(Y_t)^2=1$, so $t$ is a natural parameter for the ray. Let $\xi$ be a normalized coordinate, that is
\[
(\Psi_X)^2+(\Psi_Y)^2=1\,.
\]
It is easy to see that
\begin{equation}\label{divray}
\begin{split}
\Theta_{jXX}+\Theta_{jYY}=\frac{1}{J}\frac{d}{dt}kJ\,,
\end{split}
\end{equation}
where
\begin{equation}\label{J}
\begin{split}
J=\det \left(\begin{array}{lcr}
X_t & X_{\bar\xi} \\
Y_t & Y_{\bar\xi}
\end{array}\right)
\end{split}
\end{equation}
In the coordinates $(t,\xi)$ eq.~(\ref{MPE}) is written as
\begin{equation}\label{MPE1}
2\mathrm{i}k B_{jt} + \mathrm{i}\left(k_t+k\frac{J_t}{J}\right) B +
 B_{j\xi\xi} + \alpha_j B_j = 0\,,
\end{equation}
Substituting
\[
C_j=\sqrt{kJ}B_j\,,
\]
we get the usual non-stationary Schr\"odinger type equation
\begin{equation}\label{MPE2}
2\mathrm{i}k C_{jt} + C_{j\xi\xi} + \alpha_j C_j = 0\,,
\end{equation}
This equation can be solved effectively by many existing methods. So the main difficulty of our approach consist
 in calculating the corresponding coordinate system.
 \par
The examples of rays presented in the work \cite{Burridge} show that in this case the concentrated solutions (for which the ray-centered coordinates instead of the ray coordinates are used)  are of less importance because
the rays strongly depend on modes. Nevertheless, because the ray centered coordinates are much simpler than the ray coordinates,
the corresponding parabolic equation would be interesting. We derive such an equation in the next section.

\section{The adiabatic mode Helmholtz equation and the ray parabolic equation in ray centered coordinates}

To obtain the adiabatic mode Helmholtz equation from eq.~(\ref{MPE}), we introduce the new amplitude
\[
D_j(x,y)=B_j(X,Y,\xi)\,,
\]
where $\displaystyle{(x,y)=\frac{1}{\epsilon}(X,Y)}$ are the initial (physical) coordinates.
One can easily obtain the following formulas for the $x$-derivatives of $D_j$:
\begin{equation}\label{d1x}
D_{jx}=B_{j\xi}\cdot \sqrt{\epsilon}\Psi_X+\epsilon B_{jX}\,,
\end{equation}
\begin{equation}\label{d2x}
\begin{split}
& D_{jxx}=B_{j\xi\xi}\cdot \epsilon(\Psi_X)^2  + \epsilon^{3/2}(2B_{j\xi X}\Psi_X+B_{j\xi} \Psi_{XX})+ \epsilon^2B_{jXX}\,,
\end{split}
\end{equation}
and analogous formulas for the $y$-derivatives.
\par
The solvability condition of problem at $O(\epsilon^{3/2})$ gives us
\begin{equation*}
\begin{split}
2B_{j\xi X}\Psi_X+B_{j\xi}\Psi_{XX} + 2B_{j\xi Y}\Psi_Y+B_{j\xi} \Psi_{YY}=0\,.
\end{split}
\end{equation*}
Substituting the obtained expressions for derivatives into eq.~(\ref{MPE}) we get,
after some manipulations, the reduced Helmholtz equation for $D$
\begin{equation}\label{rhe}
\begin{split}
2\I(\theta_x D_{jx} + \theta_x D_{jy})+\I(\theta_{xx} + \theta_{yy})D_j + D_{jxx} +D_{jyy}+\bar\alpha_j D_j=0\,,
\end{split}
\end{equation}
where $\theta(x,y)=\frac{1}{\epsilon}\Theta(X,Y)$, $\bar\alpha_j=\epsilon\alpha_j$.
\par
This equation can be transformed to the usual Helmholtz equation
\begin{equation}\label{he}
\begin{split}
\bar D_{jxx} +\bar D_{jyy}+k^2 \bar D_j+\bar\alpha_j\bar D_j=0\,,
\end{split}
\end{equation}
where $k^2=(\theta_x)^2+(\theta_y)^2$ by the substitution $\bar D_j =D_j\exp(\I\theta)$.
 \par
 To obtain the ray parabolic equation in the ray-centered coordinates we first rewrite eq.~(\ref{he}) in the slow variables
 $(X,Y)=(\epsilon x,\epsilon y)$ (ray scaling)
\begin{equation}\label{hers}
\begin{split}
\epsilon^2\bar D_{jxx} +\epsilon^2\bar D_{jyy}+k^2 \bar D_j+\epsilon\alpha_j \bar D_j=0\,.
\end{split}
\end{equation}
Then, in the vicinity of a given ray, eq.~(\ref{hers}) can be written in the form
\begin{equation}\label{herc}
\begin{split}
\epsilon^2\frac{1}{h}(\frac{1}{h}\bar D_{jt})_t +\epsilon^2\frac{1}{h}(h\bar D_{jn})_n+k^2 \bar D_j+\epsilon\alpha_j \bar D_j=0\,,
\end{split}
\end{equation}
where $t$ is a natural parameter of the ray (arc length), $n$ is the (oriented) distance to the ray and
$\displaystyle{h=1-\frac{k_1}{k_0}}$. Hereafter we use, for a given function  $f=f(t,n)$, the following denotations:
$f_0=f|_{n=0}$, $f_1=f_n|_{n=0}$ and $f_2=f_{nn}|_{n=0}$.
\par
Substituting into eq.~(\ref{herc}) the Taylor expansions
\begin{equation*}\label{tayk}
\begin{split}
k^2 & =k_0^2+2k_1k_0n+(k_1^2+2k_0k_2)n^2\\ & =k_0^2+\sqrt{\epsilon}2k_1k_0N+\epsilon(k_1^2+2k_0k_2)N^2\,,
\end{split}
\end{equation*}
\begin{equation*}\label{tayk}
\begin{split}
\frac{1}{h} =1+\frac{k_1}{k_0}n+\frac{k_1^2}{k_0^2}n^2 =1+\sqrt{\epsilon}\frac{k_1}{k_0}N+\epsilon\frac{k_1^2}{k_0^2}N^2\,,
\end{split}
\end{equation*}
\begin{equation*}\label{tayk}
\begin{split}
\frac{1}{h^2} =1+2\frac{k_1}{k_0}n+3\frac{k_1^2}{k_0^2}n^2 =1+\sqrt{\epsilon}2\frac{k_1}{k_0}N+\epsilon 3\frac{k_1^2}{k_0^2}N^2\,,
\end{split}
\end{equation*}
where $\displaystyle{N=\frac{1}{\sqrt{\epsilon}}}$ (parabolic scaling), and the WKB-ansatz
$\bar D_j=(u_0+\epsilon u_1+\ldots)\exp((1/\epsilon)\theta)$, we obtain at $O(\epsilon^0)$
\begin{equation*}\label{tayk}
\begin{split}
\theta_t=k_0
\end{split}
\end{equation*}
and at $O(\epsilon^1)$ the parabolic equation in ray centered coordinates
\begin{equation*}\label{tayk}
\begin{split}
& 2\I k_0 u_{0t}+\I k_{0t}u_0+u_{0NN}
 +(2(k_0k_2-k_1^2)N^2+\alpha_{j0}) u_0=0\,.
\end{split}
\end{equation*}

 \section{Conclusion}
In this work the ray adiabatic mode parabolic equations is derived by the multiple-scale approach. In the derivation all features of shallow-water acoustics were taked into account.

\section*{Acknowledgements}
The authors are grateful for the support of "Exxon Neftegas Limited" company while undertaking this
research.


\begin {thebibliography}{99}

\bibitem{Burridge} Burridge,~R., Weinberg,~H. 1977, Horizontal rays and vertical modes. In \emph{Wave propagation and
underwater acoustics,} ed. by J.R.Keller and I.S.Papadakis, Lecture Notes in Physics, Vol.~70. Springer-Verlag, New-York.
\bibitem{collins}
Collins~M.D. 1993, The adiabatic mode parabolic equation. \emph{J. Acoust. Soc. Amer.}
V.~{\bf 94}, N.~4, pp.~2269-2278.

\bibitem{trofimov1}
Trofimov~M.Yu. 1999, Narrow-angle parabolic equations of adiabatic single-mode propagation in horizontally inhomogeneous shallow sea. \emph{Acoust. Phys.} V.~45, pp. 575-580.

\bibitem{trofimov2}
Trofimov~M.Yu. 2002,
Wide-angle mode parabolic equations. \emph{Acoust. Phys.} V.~48, pp. 728-734.

\bibitem{babich}
Babich,~V.M., Buldyrev,~V.S. 1991, Short-wavelength Diffraction Theory, Asymptotic Methods. Springer, 456 pp.


\end{thebibliography}
\end {document}